\documentclass[12pt]{article}
\begin{document}

\begin{center}
{\Large\bf
Radio Wave ``Messengers'' of Periodic Gravitational Radiation
and the Problem of Gravitationally Induced Nonlinearity
in Electrodynamic Systems
}
\vskip5mm
{\bf
A.~B.~Balakin\footnotemark[1]
{\footnotesize Corresponding Member of the RAS}
G.~V.~Kisun'ko\footnotemark[2]
and Z.~G.~Murzakhanov\footnotemark[1]
}
\vskip5mm
\end{center}

\footnotetext[1]{\it
Kazan State University, ul. Lenina
18, Kazan, 420008 Tatarstan, Russia
}
\footnotetext[2]{\it
Department of Theoretical
Problems, Russian Academy of Sciences, ul. Vesnina 12,
Moscow, 121001 Russia
}

\section*{Introduction}

Consequences of the influence of gravitational radiation on
hierarchic systems should be sought not in energy variations but
in the removal of the degeneracy with respect to latent
parameters, in revealing the latent nonequivalence of similar
subsystems, and in redistributing internal energy-information
reserves among interacting subsystems.

This statement is the core of our concept that was the basis of
the papers devoted to the study of response of electrodynamic,
kinetic, and hydrodynamic systems to gravitational radiation (see,
for example, [1--5]).

In this paper, developing this concept, we discuss a
gravitationally induced nonlinearity in hierarchic systems. We
consider the generation of extremely low-frequency radio waves
with a frequency of the periodic gravitational radiation; the
generation is due to an induced nonlinear self-action of
electromagnetic radiation in the vicinity of the
gravitational-radiation source. These radio waves are a
fundamentally new type of response of an electrodynamic system to
gravitational radiation. That is why we here use an unconventional
term: radio-wave {\it messengers} of periodic gravitational
radiation.

\section*{
Theoretical models of the gravitationally\\
induced nonlinearity in electrodynamic systems
}

Phenomenological covariant electrodynamics of dielectric media
[6, 7] is based on the evolution equations
\vskip3mm
\begin{equation}\label{1}
\nabla_k H^{ik}=0\,,\quad
\nabla_k \mathrel{\mathop{F}\limits^{*}}{}^{ik}=0\,,\quad
\mathrel{\mathop{F}\limits^{*}}{}^{ik}=\frac12\epsilon^{ikmn}F_{mn}
\end{equation}
and on the material relations
\begin{equation}\label{2}
H^{ik}=C^{ikmn}F_{mn}+S^{ikmnpq}F_{mn}F_{pq}
+D^{ikmns}\nabla_sF_{mn}+\ldots\,,
\end{equation}
which combine the Maxwellian stress tensor $F_{mn}$ and induction
tensor $H^{ik}$. Here, $\nabla_k $ is a covariant derivative,
$\epsilon^{ikmn}$ is the discriminant tensor,
and $\mathrel{\mathop{F}\limits^{*}}{}^{ik}$ is the
dual tensor of $F_{mn}$.

Tensor phenomenological coefficients $C^{ikmn}$, $S^{ikmnpq}$ and
$D^{ikmns}$ are based on the metric tensor $g_{ik}$, the Riemann
tensor $R_{ikmn}$ and its covariant derivatives, the four-vector
macroscopic velocity $U^i$ of the medium, and of the constants
characterizing dielectric and magnetic properties of the medium.
If the tensor $S^{ikmnpq}$ has nonzero components, the theory will
be beyond the scope of linear electrodynamics.

The term ``gravitationally induced nonlinearity'' emphasizes that,
for zero gravitational field, all the components of $S^{ikmnpq}$
vanish, and they become nonzero only with a curvature in the
surrounding space-time. While modeling the nonlinearity of such a
type, we obviously use the Riemann curvature tensor to construct
these coefficients. To prove theoretically the possibility that a
gravitationally induced nonlinearity exists, it is sufficient to
consider at least one bright example; therefore, we discuss the
so-called inertial-tidal nonlinearity, for which the following
relation takes place:
\begin{equation}\label{3}
H^{ik} = F^{ik}
+ QR^{ikmn} F_{ms}U^s\mathrel{\mathop{F}\limits^{*}}_{nl}U^l\,.
\end{equation}

Here, $Q$ is a phenomenological coefficient (a latent parameter of
the electrodynamic system), which is included in the solution to
electrodynamic equations (1) only with a gravitational field
($R^{ikmn}\ne0$). The terms ``tidal'' and ``inertial'' are related
to the curvature tensor and the velocity, respectively.

\section*{
Analysis of solutions
to the electrodynamic equations
}

We now consider a particular solution to set (1) under material
relations (3) for the space-time metric defining a
gravitational-radiation field:
$$
ds^2 = 2
du dv - L^2 \left(
    {\rm e}^{2\beta}{dx^2}^2
  +{\rm e}^{-2\beta}{dx^3}^2
\right)\,,
\qquad
L^{\prime\prime}(u)+\beta^{\prime2}(u)L=0\,,
$$
\begin{equation}\label{4}
u = (ct-x^1)/\sqrt2,,\quad
v = (ct+x^1)/\sqrt2,,
\end{equation}
when the electromagnetic field is described by the vector potential
$$
A_i=\delta_i^3{\rm e}^{-\beta}A(u,v)
$$
and the velocity vector has the form
\begin{equation}\label{5}
U_i = \delta_i^u U_u + \delta_i^v U_v(u) + \delta_i^2 U_2(u)\,,
\quad
U_u = (1+U_2^2{\rm e}^{-2\beta}L^{-2})/2U_v\,.
\end{equation}
In the case of dust-like matter or for a geodesic observer, the
quantifies $U_v$ and $U_2$ are constants. In such a formulation,
the problem reduces to solving one equation in the unknown
function $A(u, v)$:
\begin{equation}\label{6}
\frac{\partial^2{A}}{\partial{u}\partial{v}}
=
\frac12 Q R^3_{\cdot u3u} U_2 \frac{\partial}{\partial{v}}
\left\{
  \frac{\partial{A}}{\partial{v}}
  \left[
      U_v \frac{\partial}{\partial{u}} ({\rm e}^{-\beta} A)
    +U_u {\rm e}^{-\beta} \frac{\partial{A}}{\partial{v}}
  \right]
\right\}\,.
\end{equation}
Integrating equation (6) with respect to $v$, we obtain the key
equation
\begin{equation}\label{7}
\frac{\partial{A}}{\partial{u}}
=
\frac12 Q R^3_{\cdot u3u} U_2 {\rm e}^{-\beta}\frac{\partial{A}}{\partial{v}}
\left[
      U_u \frac{\partial{A}}{\partial{v}}
    +U_v \left( \frac{\partial{A}}{\partial{u}} - \beta^{\prime} A \right)
  \right]
+ \phi(u)\,,
\end{equation}
where $\phi(u)$ is an arbitrary function of $u$.

An ambiguity in representing the solution for an electromagnetic
wave traveling in the direction of gravitational-radiation
propagation is due to the function $\phi(u)$ [8]. This ambiguity
can be eliminated if the following conditions are set on the
hypersurface $v= 0$:
\begin{equation}\label{8}
A(u,0) = 0\,,\quad
\frac{\partial{A}}{\partial{v}}(u,0) = 0\,,
\end{equation}
and if, without gravitational radiation, the following relation is
assumed to be valid:
\begin{equation}\label{9}
A\equiv A^0(v)=E_0 (1-\cos{kv})/k\,.
\end{equation}
In view of relations (8) and (9), it follows from equation (7)
that $\phi(u)\equiv0$; the solution to equation (7) can be presented
in the form
\begin{equation}\label{10}
A(u,v) = a(u) + \sum_{n=1}^{\infty}a_n(u)\cos{nkv}
\end{equation}
We are most interested in the function $a(u)$, which does not
contain the variable $v$ and is an additive component of the
potential $A(u, v)$. If relations (8) and (9) are valid, the
derivative $a^{\prime}(u)$ becomes nonzero for $QR^3_{\cdot
u3u}U_2\ne0$. In fact, substituting (10) in (7), we find that
\begin{equation}\label{11}
a^{\prime}(u)
=
\frac14 Q R^3_{\cdot u3u} U_u U_2 k^2 {\rm e}^{-\beta}
\sum_{n=1}^{\infty}n^2a_n^2(u)\,.
\end{equation}
By virtue of relation (9), at least one coefficient $a_1$ is
nonzero; therefore, $a^{\prime}$ cannot vanish unless the factor
$Q R^3_{\cdot u3u} U_2$ is zero.

Thus, the induced nonlinear self-action of the electromagnetic
field results in, first, sequential frequency doubling and,
second, the generation of an additive component in the
electromagnetic potential. This component describes an
electromagnetic wave propagating in the reverse direction, i.e., a
reflected wave. In other words, the phase conjugation rather than
the modulation of the incident wave takes place. It is the time
dependence of the metric coefficients that governs both the
frequency and the temporal profile of the reflected wave.

Supposing the parameter $Q$ to be small ($Q\rightarrow0$) and
expanding $a_n(u)$ in a power series of Q, we obtain in the
lowest-degree approximation
\begin{equation}\label{12}
a^{\prime}(u)
=
-\frac18 Q U_2 U_v^{-1} (1 + U_2^2{\rm e}^{-2\beta}L^{-2})
(\beta^{\prime} L^2)^{\prime} L^{-2}{\rm e}^{-\beta}E_0^2\,.
\end{equation}
Therefore, $a(u)$ can be easily found in closed form. The electric
field strength of the reflected wave $E_i(u)=\delta_i^3{\rm
e}^{-\beta}(\beta^{\prime} a - a^{\prime})$ is proportional to the
energy of the incident electromagnetic field $W_0 = E_0^2 /4\pi$.
It increases with the transverse velocity of motion
($U_2\rightarrow\infty$) and also with the longitudinal velocity
($U_v\rightarrow0$). The lack of information on the magnitude of
parameter $Q$ does not allow us to directly estimate the
conjugated-wave amplitude.

\section*{Physical applications}

This exactly integrable model of the electromagnetic-field
evolution with the inertially tidal nonlinearity thus illustrates
very clearly the following idea: Periodic gravitational radiation,
which, in one way or another, initiates the nonlinear self-action
of electromagnetic waves propagating in the vicinity of the
gravitational-radiation source, generates low-frequency
electromagnetic radiation with the same frequency. This induced
radiation is not a modulation of the initial field but a new
spectral component of the electromagnetic radiation, which
announces that there is a gravitational-radiation source. These
indeed are messengers of periodic gravitational radiation. It
should be noted that the messengers of gravitational radiation may
be treated as a specific class of solutions similar to transition
radiation and transition scattering, which were introduced by
V.L.~Ginzburg and V.N.~Tsytovich [9]

In our view, the method of searching for gravitational-radiation
messengers should be as follows. The frequency and directions of
the probable gravitational radiation are known from the catalog of
astrophysical systems taken as sources of periodic gravitational
radiation. To identify these sources, a center for observation and
correlative analysis of extremely low-frequency radio waves of
cosmic origin should be established. Satellite observations appear
to be the most efficient. In addition, special ground stations,
such as the Vladimir Station for observation of extremely
low-frequency variations of the earth's natural electromagnetic
background, can significantly contribute to the solution of the
problem as well.

To find radio wave messengers of periodic gravitational radiation,
it is important not only to optimally select the
gravitational-radiation source according to its frequency,
distance, and radiation power, but to ensure that either the
source be located in a zone of generation of high-power
electromagnetic radiation or the gravitational waves traverse the
zone when traveling to the earth. From this viewpoint, the
following objects may be the most promising:

Relativistic binary pulsars in star clusters;

Solar radiation in a field of distant sources of periodic
gravitational radiation;

The earth's ionosphere in a field of distant sources of
gravitational radiation.

By describing the radio wave messengers of gravitational radiation
in each of the cases given above, we hope to discover and consider
an inherent and most efficient process of forming a
gravitationally induced nonlinearity, to carry out detailed
calculations and estimates, and to determine the optimum method of
detection.

{\hfill\it Translated by V.A. Chechin}
\end{document}